\documentclass[12pt]{iopart}
\usepackage{amsfonts,amssymb,amsfonts,textcomp}
\usepackage{iopams,cite}
\usepackage{graphicx}
\usepackage{epsfig}
\usepackage{setspace}
\usepackage{layout}

\usepackage{indentfirst}
\usepackage[T1]{fontenc}
\usepackage[latin9]{inputenc}
\usepackage{amstext}

\usepackage[english]{babel}
\usepackage{color}

\begin{document}
\title{Dynamics of relaxation to a stationary state for interacting molecular motors}
\author{Luiza V. F. Gomes and Anatoly B. Kolomeisky}
\address{Department of Chemistry and Center for Theoretical Biological Physics, Rice University, Houston, Texas, 77005, USA}
\eads{\mailto{lgf3@rice.edu}}
\eads{\mailto{tolya@rice.edu}}

\begin{abstract}
Motor proteins are active enzymatic molecules that drive a variety of biological processes, including transfer of genetic information, cellular transport, cell motility and muscles contraction. It is known that these biological molecular motors usually perform their cellular tasks by acting collectively, and there are interactions between individual motors that specify the overall collective behavior. One of the fundamental issues related to the collective dynamics of motor proteins is the question if they function at stationary-state conditions. To investigate this problem, we analyze a relaxation to the stationary state for the system of interacting molecular motors. Our approach utilizes a recently developed theoretical framework, which views the collective dynamics of motor proteins as a totally asymmetric simple exclusion process of interacting particles, where interactions are taken into account via a thermodynamically consistent approach. The dynamics of relaxation to the stationary state is analyzed using a domain-wall method that relies on a mean-field description, which takes into account some correlations. It is found that the system quickly relaxes for repulsive interactions, while attractive interactions always slow down reaching the stationary state. It is also predicted that for some range of parameters the fastest relaxation might be achieved for a weak repulsive interaction. Our theoretical predictions are tested with  Monte Carlo computer simulations. The implications of our findings for biological systems are briefly discussed.  
\end{abstract}


\newpage

\section{Introduction}

Motor proteins, also known as biological molecular motors, play important roles in supporting and maintaining various biological processes \cite{alberts_book,phillips,kolomeisky2007molecular,Chowdhury2013,kolomeisky2013motor,kolomeisky_book,mclaughlin16}. They are responsible for nucleic acids copying and repairing, cellular transport of vesicles and organelles, transfer of genetic information, synthesis of proteins and nucleic acids, muscles functioning, cell motility and signaling, and many other tasks \cite{alberts_book,phillips,kolomeisky_book}. Motor proteins act by catalyzing some specific chemical processes such as the hydrolysis of energy-rich adenosine triphosphate (ATP) or biopolymerization of nucleic acids and proteins. The released energy from these chemical reactions is then converted into a mechanical work, which supports the specific functions of the given molecular motor \cite{phillips,kolomeisky_book}. Although motor proteins have been intensively studied in the last 20 years, both experimentally and theoretically, many aspects of their mechanisms, and especially  collective dynamics properties, remain not fully understood \cite{Chowdhury2013,kolomeisky2013motor,kolomeisky_book,mclaughlin16,appert15}.

It is widely accepted that the majority of motor proteins function in  groups, and interactions between individual molecules determine the cooperative behavior of molecular motors \cite{kolomeisky_book,mclaughlin16,guerin10,walcott12,appert15}. These interactions have been measured for kinesin motor proteins, although the results are contradictory \cite{vilfan2001dynamics,roos2008dynamic,telley09}. Experiments on clustering of kinesin molecules on microtubules without the presence of the ATP molecules (energy source for the motion) suggested that these motor proteins attract each other with an interaction energy close to $1.6 \pm 0.5 \ k_{B}T$ \cite{roos2008dynamic}. At the same time, the single-molecule imaging of {\it in vitro} dynamics and processivity of kinesin molecules concluded that kinesins  most probably weakly repel during each encounter \cite{telley09}.  The importance of interactions for motor proteins stimulated multiple theoretical investigations that aimed to uncover the role of interactions in the collective dynamics \cite{campas06,slanina08,Parmeggiani2013,Pinkoviezky2013,Klumpp2004,golubeva13,vuijk15,celis2015correlations,hamid2015,midha17}. Most of them utilized totally asymmetric simple exclusion processes (TASEPs), which are  non-equilibrium multi-particle  models that have been widely employed to analyze various dynamic processes in Chemistry, Physics and Biology \cite{Derrida98,Chou11,Bressloff13}. Most of these  theoretical studies treated the effect of interactions in collective dynamics of molecular motors  in a phenomenological way \cite{campas06,slanina08,Parmeggiani2013,Pinkoviezky2013,Klumpp2004,golubeva13,vuijk15}. A more fundamental approach to describe interactions has utilized thermodynamic arguments to describe the formation and breaking contacts between neighboring motor proteins \cite{celis2015correlations,hamid2015,midha17}, providing a better microscopic connection between properties of molecular motors and their cooperative behavior. 

All existing studies on collective dynamics of motor proteins always assume that the investigated systems are in the steady state  \cite{campas06,slanina08,Parmeggiani2013,Pinkoviezky2013,Klumpp2004,golubeva13,vuijk15,celis2015correlations,hamid2015,midha17}. Although it might be a very reasonable assumption, there are no clear experimental proofs that molecular motors in cells are always acting under stationary conditions. On the contrary, based on the complex nature of cellular medium (crowding, chemical interactions with multiple particle, compartmentalization, etc.), one might suggest that the biological cells might not follow the stationary state dynamics all the time. It is critically important to understand if biological systems are always at the steady state, and if not, what is the relaxation dynamics to the stationary state conditions.

In this paper, we develop a new theoretical method to probe the  dynamics of relaxation of interacting molecular motors to their stationary state. The collective behavior of molecular motors is viewed as TASEP with  interactions that are considered using a thermodynamically consistent description \cite{celis2015correlations,hamid2015,midha17}. The process of relaxation to the steady state for the system of multiple interacting particles is analyzed using a {\it domain wall} (DW) approach \cite{DomainWall,appert15}, which reduces a complex multi-particle dynamics into an effective single-particle (domain wall) motion, in which the domain wall describes the border  between different stationary phases in the system \cite{DomainWall}.  Our theoretical calculations show that the relaxation dynamics depends on the strength of the interaction energy: it is faster for repulsions and it is slow for attractions. For some range of parameters we predict that there is the fastest relaxation dynamics, which can be achieved for the weak repulsion interactions. The implications of our theoretical predictions for biological transport are also briefly discussed.

\section{Theoretical Analysis}

\subsection{Model}

Let us view a system of interacting molecular motors moving on a cellular filament as the TASEP model of interacting particles on the lattice as illustrated in Fig. 1. This picture is inspired by the transport of kinesins along microtubule cytoskeleton filaments \cite{celis2015correlations,hamid2015,midha17}. Each lattice site can be occupied or empty, and no more than one particle can be found at the same location - this is the exclusion part of the interactions. In addition,  molecular motors can interact with each other when they are sitting on the neighboring lattice sites. The strength of this short-range interaction is assumed to be equal to $E$. The attraction corresponds to $E>0$, while the repulsions are described by negative $E$. We can associate the interaction between two neighboring particles as a creation of an effective chemical ''bond'' between them. Then it is clear that all dynamic transitions in the system can be divided in three groups. The first group of transitions does not involve changing the number of bonds, and these transitions are taking place with a rate $1$: see Fig. 1. Note here that the transition that creates one bond and simultaneously breaks another one is also taking place with the rate $1$ because there is no overall energy change. The second type of transitions is associated with creating the new bond, and they are occurring with a rate $q$ (Fig. 1). The third type of transitions leads to breaking the bond, and it happens with a rate $r$, as shown in Fig. 1. The entrance and exit to the system also depend on interactions. As illustrated in Fig. 1, if the entering particle does not make the new bond then the entrance rate is $\alpha$, while the entrance with creating the bond is associated with a rate $q \alpha$. Similarly, leaving the system without breaking the bond is given by a rate $\beta$, while the same transition with dissociating the bond is associated with a rate $r \beta$.

\begin{figure}[!ht]
  \begin{center}
    \includegraphics[width=0.9 \textwidth]{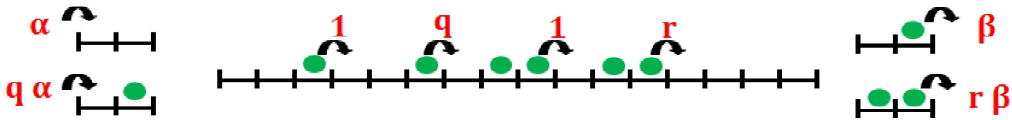}
    \caption{Schematic view of the TASEP model for interacting molecular motors. Particles on the lattice move with the rates $q$ or $r$ if the interaction bond between two neighboring particles is made or broken, respectively. In all other cases, the rate is equal to 1.  The particles enter the system with the rates $\alpha$ or $q \alpha$ when the inter-particle bond is made. The particles can leave the system with the rate $\beta$ or $r \beta $ if the inter-particle bond is broken.}
    \label{cellisfig}
  \end{center}
\end{figure}

The analogy between the inter-particle interaction and the effective chemical bond is very useful since it allows us to express the rates $q$ and $r$ \cite{celis2015correlations,hamid2015}. The detailed balance arguments suggest that
\begin{equation}
\frac{q}{r}= \exp{\left( {\frac{E}{k_{B}T}}\right)},
\end{equation}
which can be viewed as an effective equilibrium constant to create the inter-particle bond. This leads to \cite{celis2015correlations,hamid2015},
\begin{equation}\label{rates}
q=\exp{\left( {\frac{\theta E}{k_{B}T}}\right)}, \quad r=\exp{\left( {\frac{(\theta-1)E}{k_{B}T}}\right)},
\end{equation}
where a dimensionless parameter $0 \le \theta \le 1$ quantifies the effect of interactions on creating and breaking the inter-particle bond.  For the limiting case $\theta=0$, the bond formation rate $q$ is independent of the interaction energy, while the bond breaking rate $r$ strongly depends on it. For $\theta=1$ the trend is reversed: the bond formation depends on the interaction between particles, while the dissociating the bond is independent of the interaction. For other  values of $\theta$, $0 <\theta <1$, which seems to be a more realistic situation, both transitions are modified by interactions as specified by Eq. (\ref{rates}).

We can also explain the physical meaning of the transitions rates $q$ and $r$ given in Eq. (\ref{rates}). If the formation of the bond is energetically favorable ($E>0$), then the transition to this configuration is faster, $q>1$, while leaving this configuration is slower, $r<1$. For the repulsive inter-particle interactions ($E<0$), it is actually faster to break the bond ($r>1$), and it is slower to create a new one ($q<1$). In the situation without interactions, $E=0$, the bond association and dissociation transitions have the same speeds, $q=r=1$, and the system is identical to the well-studied TASEP without interactions and with open open boundary conditions \cite{Derrida98,Chou11,Derrida}.

\subsection{Mean-field theory with correlations}

To understand the relaxation dynamics we first develop a theoretical description for the stationary state conditions. It has been shown before that simple mean field treatments that completely neglect the correlations in the system lead to unphysical results for TASEP with interactions \cite{hamid2015}. For this reason, we utilize and extend the approach that takes into account some correlations, providing a much better  description of the dynamics in the system \cite{celis2015correlations}.

The main idea of this method is to analyze clusters of two neighboring sites \cite{celis2015correlations}. Depending on their occupancy, each of them can be found in one of four possible states: we label them as $(0,0)$ for the fully empty cluster, $(1,0)$ or $(0,1)$ for the half-occupied clusters, and $(1,1)$ for the fully occupied cluster. The corresponding stationary-state probabilities of these states are defined by $P_{00}$, $P_{10}$, $P_{01}$ and $P_{11}$, respectively. The conservation of probability requires that
\begin{equation}\label{prob_conservation}
P_{00}+P_{10}+P_{01}+P_{11}=1.
\end{equation}
We can also connect these probabilities to the particle density $\rho$, which is assumed to be constant in the bulk of the system,
\begin{equation}\label{rho}
P_{10}+P_{11}=\rho, \quad P_{01}+P_{11}=\rho.
\end{equation}

To calculate the fluxes, one has to consider four lattice sites segments that describe all situations with non-zero fluxes if the middle cluster is always in the state $(1,0)$ \cite{celis2015correlations}. They are presented in Fig. 2. Then the overall bulk current, $J_{b}$, can be written as a sum four fluxes for each segment,
\begin{equation}
J_{b}=J_{1}+J_{2}+J_{3}+J_{4},
\end{equation}
and the explicit expressions for fluxes are given by
\begin{equation}\label{J1}
J_{1}=\gamma P_{10} \left(\frac{P_{00}}{P_{00}+P_{01}}\right),
\end{equation}
\begin{equation}\label{J2}
J_{2}=(1-\gamma) r P_{10}\left(\frac{P_{00}}{P_{00}+P_{01}}\right),
\end{equation} 
\begin{equation}\label{J3}
J_{3}=\gamma q P_{10}\left(\frac{P_{01}}{P_{00}+P_{01}}\right),
\end{equation} 
\begin{equation}\label{J4}
J_{4}=(1-\gamma)  P_{10}\left(\frac{P_{01}}{P_{00}+P_{01}}\right),
\end{equation}
where $\gamma = 1/[1+\exp{(E/k_{B}T)}]$. Let us explain the physical meaning of these expressions using as an example the current from the second configuration in Fig. 2, $J_{2}$. It is a product of four terms and it describes the transition of the particle from the second site to the third site: see Fig. 2. The first term, $(1-\gamma)=\exp{(E/k_{B}T)}/[1+\exp{(E/k_{B}T)}]$, is the probability to have the particle at the first site of the segment, assuming a local equilibrium for  a two-state process (the occupied state with the energy $E$ due top the presence of the particle at the second site of the segment, or the empty state  with zero energy). In simple terms, this is just a Boltzmann's factor for this two-state process. The second term,  is the rate of the particle transition in this configuration, which is equal to $r$ because of the breaking the bond between particles. The third term is the probability that the middle cluster is in the conformation $(1,0)$. The last term gives the relative probability to have the last two sites not occupied. Because in the middle of the segment we always have the configuration $(1,0)$, the last two sites can only be in states $(0,0)$ or $(0,1)$.  All other fluxes can be explained using similar arguments. This procedure effectively allows us to take into account the correlations in the system, although relatively short in range. It is important to note that the presented method slightly differs from the one developed earlier \cite{celis2015correlations} due to the improved description of the correlations in the last two sites of the segments.

\begin{figure}[!ht]
  \begin{center}
    \includegraphics[width=0.4 \textwidth]{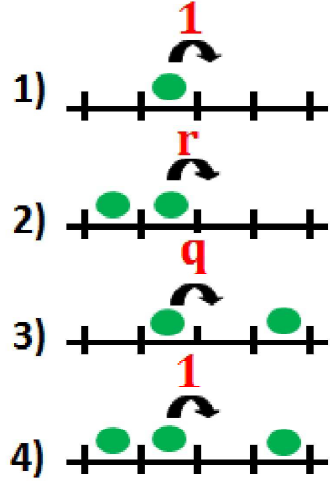}
    \caption{Four lattice segments that lead to the flux along the lattice in TASEP with interactions.}
    \label{foursites}
  \end{center}
\end{figure}

To calculate explicitly the stationary-state properties of the system, an additional approximation for the function $P_{10}$ needs to be introduced, and it can be shown that \cite{celis2015correlations}
\begin{equation}\label{P10rho}
P_{10} \simeq \frac{\rho(1-\rho)}{1 - \rho + \rho \exp{\left(\frac{E}{k_{B}T}\right)}}.
\end{equation}
This simply reflects the fact that if the cluster would have two sites occupied its probability will be modified by the interaction via the usual Boltzmann's factor \cite{celis2015correlations}. Then $(1-\rho)/\left[1 - \rho + \rho \exp{\left(\frac{E}{k_{B}T}\right)}\right]$ is the relative probability to have the second site empty if the first one is occupied.

Now combining Eqs. (3)-(10), we can obtain the expression for the bulk current,
\begin{equation}\label{current_mc}
 J_b = \frac{A \rho(1-\rho)\left[1-2\rho +\rho \exp{\left(\frac{E}{k_{B}T}\right)}\right] +B \rho^2 (1 - \rho)}{\left[1-\rho+\rho \exp{\left(\frac{E}{k_{B}T}\right)}\right]^2},
\end{equation}
with $A=[\gamma+r(1-\gamma)]$ and $B=[1+\gamma(q-1)]$. At the boundaries of the system, entrance and exit rates specify the dynamics. It was already shown that the entrance and exit currents for the TASEP with interactions are given by  \cite{celis2015correlations}
\begin{equation}\label{current_ld}
J_{entr} = \frac{\alpha(1-\rho)\left[1-2\rho+\rho\exp{\left(\frac{E}{k_{B}T}\right)}\right]+\alpha q\rho(1-\rho)}{1 - \rho + \rho \exp{\left(\frac{E}{k_{B}T}\right)}},
\end{equation}
and
\begin{equation}\label{current_hd}
J_{exit} = \frac{\beta \rho\left [1-\rho + r\rho \exp{\left(\frac{E}{k_{B}T}\right)}\right]}{1 - \rho + \rho \exp{\left(\frac{E}{k_{B}T}\right)}}.
\end{equation}

Analysis of  Eqs. (\ref{current_mc}), (\ref{current_ld}) and (\ref{current_hd}) suggests that, similarly to original TASEP without interactions, there are three stationary phases in this system. If the dynamics in the bulk is the rate limiting step then the system can be found in a maximal-current (MC) phase with the current given by Eq. (\ref{current_mc}), and the bulk density can be found from the condition, $\frac{\partial J_{b}}{\partial \rho}=0$, which leads to  
\begin{eqnarray}
& & \left[A\left(\exp{\left(\frac{2E}{k_{B}T}\right)}-3\exp{\left(\frac{E}{k_{B}T}\right)}+2\right) +B\left(\exp{\left(\frac{E}{k_{B}T}\right)}-1\right)\right] \rho^{3}  \nonumber \\
& & + \left[ 3A \left(\exp{\left(\frac{E}{k_{B}T}\right)}-2\right)+3B\right] \rho^{2}-\left[A\left(\exp{\left(\frac{E}{k_{B}T}\right)}-5\right)+2B \right]\rho \nonumber \\
& &-A=0. 
\end{eqnarray}
The bulk density in the MC phase is found by solving this cubic equation and choosing the physically reasonable root ($0 \le \rho \le 1$). If  entrance controls the stationary dynamics in the system, we have a low-density (LD) phase with the current given by Eq. (\ref{current_ld}). The connection between the entrance rate $\alpha$ and the particle density $\rho$ can be obtained  using the condition $J_{entr}=J_{b}$, yielding
\begin{equation}\label{alpha}
\alpha = \frac{A \rho \left[1-2\rho +\rho \exp{\left(\frac{E}{k_{B}T}\right)}\right] +B \rho^2}{\left[1-\rho+\rho \exp{\left(\frac{E}{k_{B}T}\right)}\right]\left[1 -2\rho +\rho \exp{\left(\frac{E}{k_{B}T}\right)} +  q \rho\right]}.
\end{equation}
Solving this equation will give the bulk density $\rho_{LD}$ in the LD phase in terms of $\alpha$ and the interaction energy $E$. Similar analysis can be done for a high-density (HD) phase where the exiting the system is the rate-limiting step. The current in the HD phase is given by Eq. (\ref{current_hd}), and the exit rate $\beta$ is related to the particle density via $J_{exit}=J_{b}$, producing 
\begin{equation}\label{beta}
\beta = \frac{A (1-\rho)\left[1-2\rho +\rho \exp{\left(\frac{E}{k_{B}T}\right)}\right] +B \rho (1 - \rho)}{\left[1-\rho+\rho\exp{\left(\frac{E}{k_{B}T}\right)}\right]\left[1-\rho+r\rho \exp{\left(\frac{E}{k_{B}T}\right)}\right]}.
\end{equation}
From this equation, one can easily obtain the bulk particle density $\rho_{HD}$ in the HD phase in terms of $\beta$ and the interaction energy $E$.

This mean-field analysis provides a satisfactory description of the TASEP with interactions, and some of these results are shown in Figs. 3 and 4. Theoretical predictions for current and particle densities in the MC phase agree quite well with computer Monte Carlo simulations for repulsions ($E<0$), while for attractive interactions this theory correctly identifies only the trend for the current and it is not so successful in predicting the bulk densities in MC phase. Similar results are found on other dynamic phases. These observations have been understood by taking into account the correlations in the system \cite{celis2015correlations}. The presented mean-field method takes into account some correlations, and this is enough to describe the repulsions where such correlations indeed are relatively short. However, for attractive interactions, the correlations are long ranged, and the theory is not successful.

\begin{figure}[h]
\centering
\includegraphics[clip,width=0.65\textwidth]{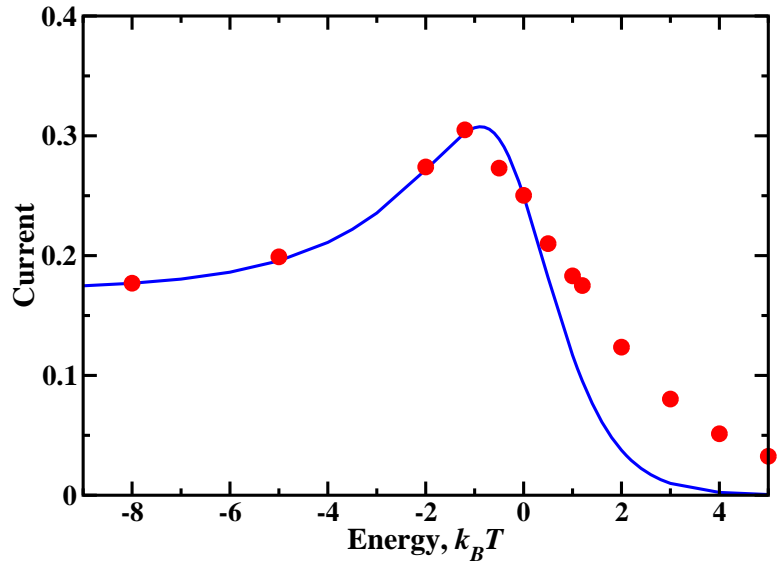}
\caption{Particle current in the MC phase as a function of the interaction energy. The curve shows theoretical predictions, while symbols correspond to Monte Carlo simulations. The following parameters were utilized: $\alpha=\beta=1$, $\theta=0.5$.}
\end{figure}

\begin{figure}[h]
\centering
\includegraphics[clip,width=0.65\textwidth]{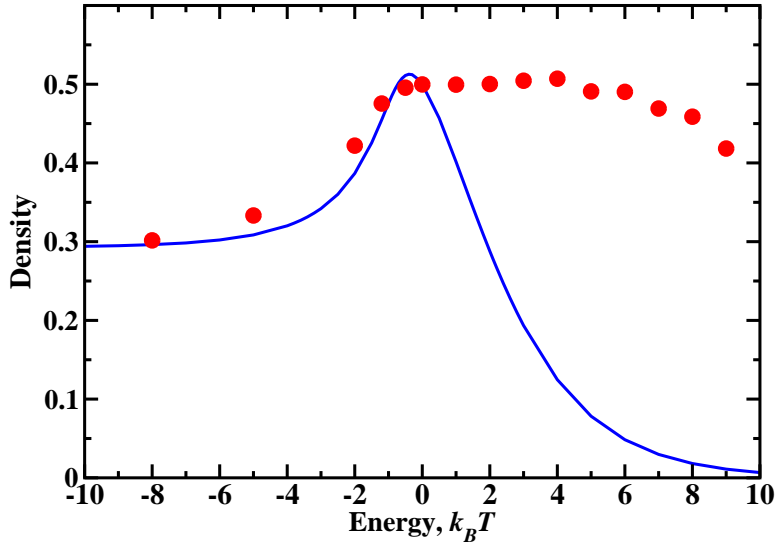}
\caption{Particle density in the MC phase as a function of the interaction energy. The curve shows theoretical predictions, while symbols correspond to Monte Carlo simulations. The following parameters were utilized: $\alpha=\beta=1$, $\theta=0.5$.}
\end{figure}

\subsection{Domain Wall Approach and Relaxation to the Stationary State}

A domain wall (DW)  approach is a method of describing dynamic phenomena in non-equilibrium low-dimensional systems by reducing multi-particle motions into a dynamics of a single ''effective'' particle, which is called the domain wall \cite{DomainWall,appert15}. This is a powerful theoretical method that provides intuitive physical explanations for the mechanisms of complex processes that are taking place in these non-equilibrium systems. There are many advantages of this approach, including simple formalism, easy application to a large number of exclusion processes, the successful application of the method for the finite-size systems, and, what is the most important for us, the ability to describe well the non-stationary processes in asymmetric exclusion processes \cite{appert15}.  

The ideas behind the DW approach are quite simple \cite{DomainWall}. It can be explained in the following way. Each of the boundaries is trying to enforce its own stationary phase in the system, and the domain wall is the border  between these two stationary segments. Then any transition in the system, such as hoping along the lattice, entering or exiting,  can be associated with the random walk of the domain wall. Depending on the parameters, one or another stationary phase eventually wins, which means that the motion of the DW is biased in the specified direction. For example, if the LD phase is the stationary phase in the system, the DW will move and fluctuate near the exit. Similarly, if the HD phase is dominating, the DW wall will be found near the entrance to the system. In the case when both phases are equally probable (dynamic phase transition), the DW performs unbiased random walk and it can be found with equal probability anywhere on the lattice. These simple arguments fully describe stationary dynamics in TASEP systems, providing a physically clear picture of underlying processes  \cite{DomainWall,appert15}.

To make the method more quantitative, let us consider, for simplicity, the DW between HD and LD phases. Then using the continuity equation, it can be shown that the velocity of the DW is given by \cite{DomainWall,appert15}
\begin{equation}
V=\frac{J_{HD}-J_{LD}}{\rho_{HD}-\rho_{LD}},
\end{equation}
where $J_{HD}=J_{exit}$ and $J_{LD}=J_{entr}$. Since the DW motion can be viewed as a biased random walk, we can define forward and backward hopping rates of the DW as $u$ and $w$, respectively. These hopping rates must be related to the overall velocity of the DW as \cite{appert15}
\begin{equation}
V=u-w.
\end{equation} 
For TASEPs with random sequential updates it was argued that the hoping rates can be written explicitly as
\begin{equation}
u=\frac{J_{HD}}{\rho_{HD}-\rho_{LD}}, \quad w=\frac{J_{LD}}{\rho_{HD}-\rho_{LD}}.
\end{equation}
One can also define then a diffusion constant of the DW, which is given by
\begin{equation}\label{diff}
D=\frac{u+w}{2}=\frac{J_{HD}+J_{LD}}{2(\rho_{HD}-\rho_{LD})}.
\end{equation}

Our goal is to utilize the DW approach to calculate the relaxation time for any arbitrary TASEP with interactions to return to its stationary state. The idea here is that the relaxation speed depends on the mobility of particles. The larger the mobility, the faster the system will return to the steady-state conditions because it is able to quickly explore more regions in the phase space. Instead of looking at the mobilities of all particles in the system, we can follow the mobility of the DW, which is specified by its diffusion constant. Thus, we suggest that the relaxation time to the stationary state  is inversely proportional to the diffusion constant of the DW,
\begin{equation}
T \sim 1/D.
\end{equation}

\begin{figure}[h]
\centering
\includegraphics[clip,width=0.65\textwidth]{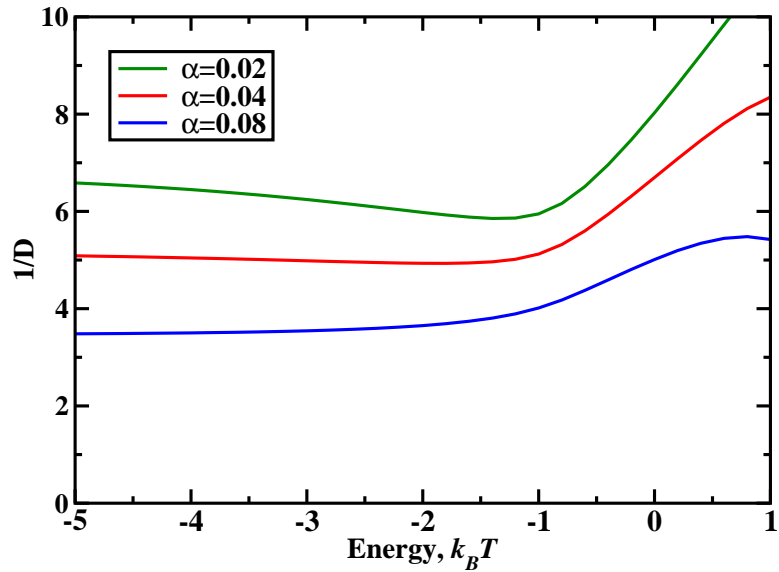}
\caption{Inverse diffusion constant of the DW, which is a measure of the relaxation dynamics to the stationary state, as a function of the interaction energy for the exit rate $\beta=0.1$ and for different entrance rates $\alpha$. }
\end{figure}

\begin{figure}[h]
\centering
\includegraphics[clip,width=0.65\textwidth]{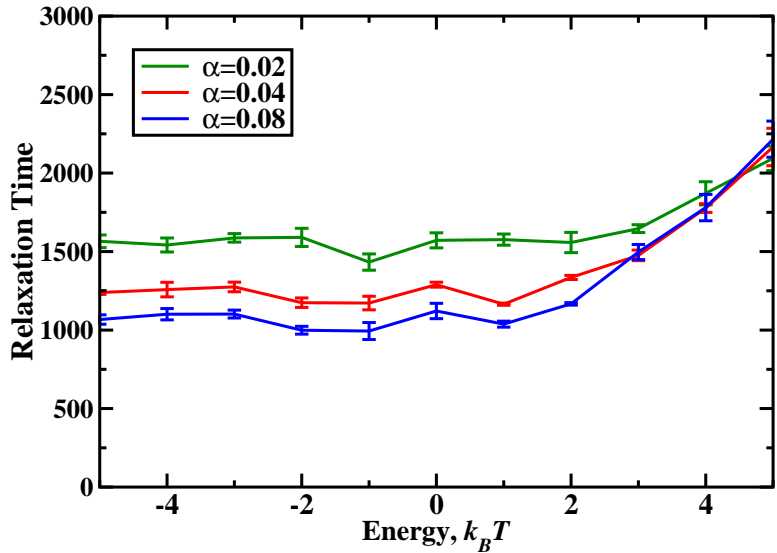}
\caption{Computer simulations of the relaxation times, measured in numbers of Monte Carlo steps, to the stationary state as a function of the interaction strength. The system was started completely empty and the following parameters were utilized for calculations: $L=1000$ and $\beta=0.1$.}
\end{figure}

To estimate the relaxation times in the TASEP with interactions, we can now apply the DW method. It is important to note that the DW approach relies on the proper  mean-field description of the system, and we already  developed such theory, as explained above. Our calculations proceed in the following way. For given values of the entrance rate $\alpha$, exit rate $\beta$ and the interaction energy $E$, first the bulk particle densities $\rho_{LD}$ and $\rho_{HD}$ are obtained from Eqs. (\ref{alpha}) and (\ref{beta}). Then the particle currents $J_{LD}$ and $J_{HD}$ are calculated using Eqs. (\ref{current_ld}) and (\ref{current_hd}). Finally, it allows us to evaluate the diffusion constant of the DW via Eq. (\ref{diff}), and the relaxation times are assumed to be proportional to $1/D$. The results of theoretical calculations are presented in Fig. 5. We predict that for the repulsive interactions the relaxation dynamics is almost independent on the strength of interactions. Relaxation times start to increase for attractive interactions. One can also see that there is a range of parameters when the most optimal relaxation might be achieved for weak repulsions (the upper curve in Fig. 5), although the minimum is not deep. Intriguingly, the fastest relaxation is predicted to be observed at the approximately similar weak repulsion strength that leads to the maximal current in the system: compare with Fig 3. In addition, we can see that for the fixed exit rate increasing the entrance rates $\alpha$ shortens the relaxation times.  

These observations can be understood if we recall the nature of transitions in TASEP with interactions. For attractions, the system tends to cluster particles together because it is energetically more favorable to make inter-particle bonds. It is hard to break single particles away in this case. This trapping of particles would significantly slow down the mobility of the system. As the result, the relaxation to the stationary state is slow.  In contrast, for repulsions, particles are not trapped (energetically not favorable to have clusters), and the interactions can even push particles forward faster and break the existing clusters. This should definitely increase the mobility, accelerating the relaxation dynamics to the steady-state conditions. At the same time, the repulsions also have a slight negative effect on relaxation by lowering the number of particles in the system, but for the weak repulsions probably this effect is not essential. It is clear that this negative effect would be the strongest for weak entrance rates, as one can see in Fig. 5. Furthermore, the increase in the entrance rates $\alpha$ puts more particles to the system, allowing for more explorations of the configurational space and leading faster to the stationary state.

Monte Carlo computer simulations, as illustrated in Fig. 6, were utilized to test our theoretical predictions. Although the data are quite noisy, one can see that most predictions of our theoretical description for the relaxation are confirmed. The fastest dynamics is observed for repulsive interactions, while the system slows down significantly for attractive interactions. Increasing the input of particles into the system also makes faster reaching the stationary state. At the same time, there are differences between theoretical predictions and computer simulations. Although, there is a minimum in the upper curve in Fig. 6  around $E=-1$ $k_{B}T$, similarly to theoretical predictions in Fig. 5, it is not clear if this result is real because of error bars and large fluctuations for all simulations curves. In addition,  computer simulations suggest that the relaxation dynamics slows down only for relatively strong attractive interactions (larger than $E \simeq 2-3$ $k_{B}T$), while our theory predicts the increase in relaxation times starting  from already  $E \simeq -0.5$ $k_{B}T$. 

It is interesting also to discuss the importance of our results for biological molecular motors. It is suggested that motor proteins that repel each other can return to the stationary-state conditions  faster than the non-interacting or motors with the positive interactions. This might be beneficiary for biological processes by making them more robust with respect to external perturbations. We predict that motor proteins with weak repulsive interactions would have the most optimal performance in the cellular transport by supporting the largest fluxes and the fastest relaxation to the stationary state. Testing this prediction in experiments might clarify better the mechanisms of motor proteins.

\section{Summary and Conclusions}

We developed a theoretical method to analyze the relaxation dynamics of interacting molecular motors to their stationary state. Since the motor proteins are moving along linear filaments, the process was mapped into totally asymmetric simple exclusion process with interactions. To investigate the process of returning to the steady-state conditions, the domain-wall method was employed. This method approximates the multi-particle dynamics in the system by a random walk of the new effective particle that coincides with the border between different stationary phases in the system. Because the DW approach relies on the successful mean-field description of the system, we modified the existing analysis of the TASEP model with interactions to take into account better the correlations. This allowed us to estimate the stationary-state properties of the system, which agree well with computer simulations. We evaluated then the mobility of the DW by calculating its diffusion constant for various ranges of parameters. It has been argued that the relaxation times are inversely proportional to these diffusion constants. This approach allowed us to quantify the role of the interactions in the relaxation to the steady-state conditions for interacting molecular motors. Our calculations show that repulsions lead to fast relaxation, while attractions slow down the return to the stationary state. Our theory predicts that for some range of parameters the most optimal relaxation dynamics might be achieved for weak repulsions. These observations are explained in terms of the formation and breaking particle clusters, and the changes in particle density in the system. Monte Carlo computer simulations mostly support our theoretical conclusions. We also argue that from the biological point of view the weak repulsions between motor protein might be beneficiary by making the biological systems more robust to external perturbations.

Although our theoretical method provides some physical insight on the mechanisms of interacting molecular motors, it is crucial to note its limitations. The presented theoretical model is rather oversimplified. It neglects the possibility of associations/dissociations at every site and the multi-filament nature of the protein tracks, which are real features of motor proteins in cellular transport. Furthermore, the application of the DW  method is limited to low-density and high-density regimes. It will be important to analyze these phenomena using more advanced theoretical methods.

\section*{Acknowledgments}
The work was supported by the Welch Foundation (Grant C-1559), by the NSF (Grant CHE-1360979), and by the Center for Theoretical Biological Physics sponsored by the NSF (Grant PHY-1427654). It was also carried out with the support of CNPq, Conselho Nacional de Desenvolvimento Científico e Tecnológico - Brasil.

\end{document}